# Max Weinstein: Physics, Philosophy, Pandeism

Helge Kragh[*]

**Abstract**: This is a brief introduction to the life and ideas of the Lithuanian-German physicist, philosopher and religious thinker Max B. Weinstein, who is today best known for his thoughts concerning so-called pandeism. An accomplished theoretical physicist and characteristic figure in the *fin-de-siècle* transition of the physical world view, Weinstein contributed to a wide range of physics. He was an early if not unsympathetic critic of Einstein's theory of relativity and Minkowski's formulation of it. In this respect he was unoriginal, but in the broader context of science and humanist culture he was an original thinker. He argued in favour of what he called pandeism, a religious-philosophical view according to which the physical world and the cosmic deity are one and the same.

## 1. Introduction

Max Bernhard Weinstein was born 1 September 1852 in Kaunas, Lithuania, by Jewish parents at a time when Lithuania was under Russian rule. He died in Berlin on 25 March 1918. After been adopted by an uncle, in 1865 Weinstein attended a Gymnasium in Insterburg (now Chernyakhovsk in the Kaliningrad region) from where he graduated in 1874. He subsequently matriculated to the University of Breslau, soon to move on to the Friedrich-Wilhelm University in Berlin. In 1878 he was naturalized as a citizen of Germany. For details on Weinstein's life and career, see [Starikov 2019, forthcoming].

Weinstein spent most of his life and career in Berlin, which at the end of the nineteenth century was not only the capital of Germany but also the capital of international physics. In 1880 he was awarded the doctoral degree which opened up for a position as untenured lecturer (*Privatdozent*) at the famous University of Berlin. Six years later he passed his *Habilitation* exam under Helmholtz allowing him to teach at the university. During a period of more than three decades he taught courses

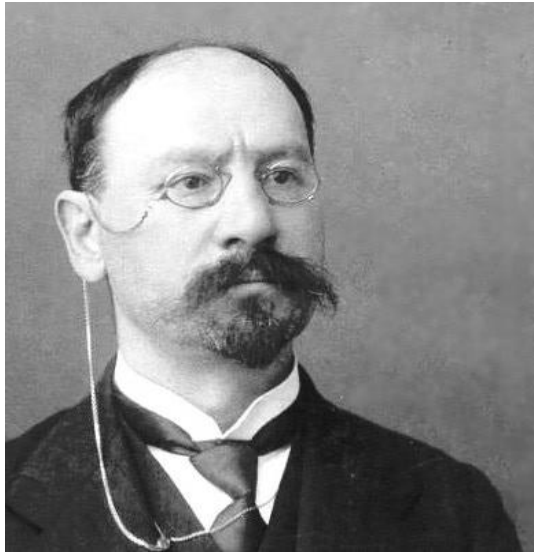

Max Weinstein, photography of 1910.

in physics and geography while at the same time contributing to the research literature in physics and its allied sciences. Apart from several monographs and chapters in recognized handbooks he wrote a series of research papers in *Annalen der Physik*, *Zeitschrift für physikalische Chemie*, *Elektrotechnische Zeitschrift* and other journals. The topics of these papers covered a broad range of areas including thermodynamics, capillarity, electrochemistry, electromagnetism, solid-state physics, meteorology, and terrestrial magnetism [Poggendorff 1898-1926; Starikov 2018].

Although Weinstein's research interests spanned experimental as well as theoretical physics, he was particularly occupied with the latter aspects such as represented by the theories of classical mechanics and electrodynamics. Still in the early 1880s J. C. Maxwell's fundamental field theory of electromagnetism was little known and appreciated in German physics, where it faced opposition by rival theories based on directly interacting electrical particles. When Weinstein in 1883 made an authorized German translation of Maxwell's important *Treatise on Electricity and Magnetism* from 1865, it made the theory better known in Germany, where it soon was generally accepted and paved the way for the electron theory [Maxwell 1883]. Of more historical than scientific interest, the following year he prepared a German translation of Fourier's classical treatise *Théorique Analytique de la Chaleur* dating from 1822. As yet another example of Weinstein's interest and competence in theoretical physics may be mentioned an introductory textbook on advanced mathematical physics [Weinstein 1901].

## 2. Relativity physics

When Weinstein turns up in the history of physics it is most often in footnotes and as a peripheral figure in connection with the early response to Einstein's theory of relativity [e.g., Pyenson 1987; Crelinstein 2006]. By heart and training a classical physicist he responded critically if not dismissively to Einstein's theory, which around 1910 attracted much attention in the German physics community. Although he understood the theory from a technical point of view, he doubted that it represented a valid physical description of nature. For example, like many other physicists at the time he was unwilling to abandon the concept of the ether as a foundation of physics. In monographs and articles between 1911 and 1914 he spelled out his objections and, at the same time, provided useful summaries of the theories of Einstein and Minkowski [Weinstein 1913b; Weinstein 1914a; Weinstein 1914b]. His book of 1913 was reviewed by Max Born [1914], a rising star in German theoretical physics and a specialist in relativistic mechanics. Apart from Max von Laue's *Das Relativitätsprinzip* from 1911, Weinstein's book was the first monograph covering the theory of relativity [see Goenner 1988].

Weinstein's objections to relativity theory were in part conceptual and in part mathematical. In 1911 he denied that quantities such as length, time and mass were relative in a physical sense, arguing that the mathematical constructions of the theory of relativity were illegitimately interpreted in physical terms. "In a nutshell, no physical implications follow from this theory, but only formal-geometric implications" [Weinstein 1911, p. 156]. He found it hard to believe that whereas in Einstein's theory a moving sphere of matter appears as a flattened ellipsoid, the form of a spherical light sphere is invariant. Although Weinstein's own exposition of relativity theory was an orgy of mathematical manipulations, he was worried of the "mystical symbolism" that threatened to replace physical reality with an unintelligible mathematical formalism [Pyenson 1985, p. 150]. Concerning the meaning of the relativity principle, Weinstein [1914c, p. 1] expressed the view that it "has often been extended so far and without reflection that an intolerable intolerance of other opinions has developed to the most silly assertions, which is almost comparable to a medieval constraint of belief."

Although Weinstein did not accept the theory as a physically valid description of nature, on the other hand he [1914b, p. 929] admitted that "Einstein's introduction of the theory of relativity will always count as one of the most important achievements in science ever." His rather critical attitude to Einstein's theory was by no means exceptional as the theory was widely considered controversial not least because of the privileged status it ascribed to light signals. More concretely, the main point of Weinstein's informed critique was that the valid deductions from the relativity principle were only concerned with "the mathematical form of the course of events, but not with the events themselves." From this and other objections he concluded that it was "questionable whether the relativity principle could be justified also in some purely physical sense" [Weinstein 1913b, p. 341; see the translated excerpts in Starikov 2015].

Although there is no indication that Weinstein and Einstein ever met, the latter was aware of Weinstein's work. In a letter of March 1916 the astronomer Wilhelm Förster called attention to a newspaper article in which Weinstein [1916a] reflected on Einstein's new theory of gravitation as expounded in his general theory of relativity. According to Weinstein, general relativity had turned gravity into a "world power" that controlled all laws of physics and necessitated a revision of physics as well as mathematics. Referring to the anxiety that Einstein's theory had caused in parts of the German populace, Förster mentioned that it was perhaps from "some recent popular explanations by our colleague Weinstein … that this unease was amplified and disseminated" [Einstein 1998, p. 275].

As regards the mathematical language of relativity theory Weinstein focused as much on Minkowski's innovations as those of Einstein. He tended to consider Einstein's special theory of relativity to be merely a particular case of Minkowski's more fundamental theory. On the one hand, he was concerned over Minkowski's mathematical formalism, which he found was "unspeakably difficult to understand." On the other hand, he dedicated his book of 1913 to Minkowski, who had passed away four years earlier, and praised him as "a man of genius" [Weinstein 1913, p. vi; Walter 1999, p. 104; Weinstein 1910b, p. 929]. Weinstein was sceptical with respect to the use of complex quantities in physics because imaginary terms could not correspond to something observable. He consequently preferred to write the formulae of relativity theory in terms of either three-dimensional vectors or components, which to his mind made them more comprehensible.

Despite his dislike of complex four-dimensional quantities, Weinstein considered Minkowski's use of them to be "one of the greatest revolutions in our accepted views" [Weinstein 1913b, p. vi]. Indeed, he was sceptical with respect to the theory of relativity, but he was neither an anti-relativist nor merely a "populariser" of the theory [Crelinstein 2006, p. 102]. In some respects his attitude to relativity was similar to the one held by other sceptics such as Emil Wiechert and Max Abraham, who mastered the theory's mathematical apparatus without accepting it as physically true. It was only after World War I that Einstein's theory became truly controversial, attracting massive criticism from scientists, philosophers and the general public in Germany as well as abroad. By then Weinstein had passed away and his earlier objections were largely if not completely forgotten. Thus, he was *post mortem* included as an anti-Einstein author in a notorious anthology with the telling title "One Hundred Authors against Einstein" [Israel, Ruckhaber and Weinmann 1931, pp. 100-101; for this work, see Goenner 1994].

## 3. Other works in physics

In many respects Weinstein was an average German *fin-de-siècle* physicist, an able researcher but of no particular originality and in no way comparable to giants such as Max Planck, H. A. Lorentz and Ludwig Boltzmann. Nonetheless, he had a solid reputation and was, for example, invited to contribute to the festschrift published on the occasion of Boltzmann's 60-year's birthday. The subject of his contribution was entropy changes in viscous physical systems, a topic he also dealt with in later publications [Weinstein 1904; Weinstein 1916b].

As seen in retrospect, his most important scientific works were within areas such as solid-state physics and physical chemistry, where he contributed to frontier research by making advanced use of quantum theory, thermodynamics and statistical mechanics. His work in this wide area included four volumes on the thermodynamics and kinetics of substances published between 1901 and 1908 [Weinstein 1901-1908]. The first volume was reviewed by the noted experimental physicist Ernst Pringsheim [1902], who found it to be generally valuable but unsuited as a textbook. One of Weinstein's papers was a critical examination of Walther Nernst's heat theorem also known as the third law of thermodynamics. In this paper Weinstein [1917] commented on the status of Nernst's theorem as compared to Einstein's theory of relativity:

> It is better to designate Nernst's theorem as a law of approximation … which ultimately all laws of nature arguably merely are – with the possible exception of the law of relativity, should it turn out to be true – than to burden it with a prediction, whose truth already now is being put into more than doubt by experience.

According to Evgeni Starikov [2018], Weinstein was an "outstanding theoretical physicist" who contributed significantly to the foundation of thermodynamics and statistical mechanics.

## 4. Philosophical views

Weinstein lived at a time when science, art and culture were to some extent in a state of turmoil characterized by a zeitgeist favouring alternatives to the traditional mechanical world view. The *fin-de-siècle* period from about 1890 to 1910 is in retrospect the revolutionary phase during which the new theories of relativity and quanta changed the foundation of physics but from a contemporary point of view, things looked different [Kragh 2014]. While Weinstein was relatively undistinguished as a physicist, he was an original thinker in the broader context of science and humanist culture, perhaps a natural philosopher rather than a physicist in the more restricted meaning of the term. He clearly had an unusually broad and deep knowledge of philosophical streams of thought through history, a desire to understand the world in the sense of *Weltanschauung* (world view) and not merely as a complex mechanism describable in terms of mathematically formulated laws of physics.

In 1904 Weinstein was appointed titular professor in philosophy and over the following years he gave a series of lectures on natural philosophy and the philosophical foundation of science to students at the University of Berlin [Gerhardt, Mehring and Rindet 1999, p. 108]. The published version of the lecture series, a comprehensive volume of more than 500 pages, indicates the breadth of Weinstein's philosophical interests and how different his views were compared to the predominant philosophy of science of a more positivistic orientation. For example, he spent three lectures dealing with the soul and spiritual capacities. Other of his lectures dealt with perception psychology, epistemology, idealism versus realism, consciousness, substantiality, infinity, and causality. While philosophers of the naturalist and positivist schools tended to disregard the human soul or deny its existence, according to Weinstein the soul was all-important and no less real than immaterial physical forces such as electricity, magnetism and gravity.

At the end of his book Weinstein [1906, p. 528] emphasized subjectivism over objectivism: "Truly, ultimately man lives only in his inner world; what man put forth from himself into the outer world is his own proper life-content." Coming from the pen of a physicist, this is a remarkable claim. It may call in mind the later views of Arthur Eddington [1920, p. 200] according to whom, in advanced science "the mind has but regained from nature that which the mind has put into nature."

Weinstein's book on natural philosophy was briefly reviewed by the eminent physical chemist Wilhelm Ostwald [1908] in *Annalen der Naturphilosophie*, a journal he had founded in 1901. In another review, which was longer and generally positive, the American philosopher Wilmon H. Sheldon pointed out the subjective character of the author's philosophy and how it thoroughly denied the mechanical view of the soul. According to Sheldon [1908], "The author regards the soul as a good concept … proof of its existence lies only in its inner conviction."

### 5. Cosmological speculations

Among the many issues in *Die philosophischen Grundlagen* was also the much-discussed question of the so-called heat death (*Wärmetod*), which Weinstein dealt with in the 29th lecture. To put it briefly, although the energy of a closed system is conserved, according to the second law of thermodynamics the free or useful energy decreases continually or, in the formulation due to Rudolf Clausius and dating from 1868, the entropy of the system increases irreversibly. Since the entropy is a measure of the system's lack of order or lack of degree of organization, the second law implies that any closed system will eventually reach a state of thermal equilibrium in which life and other forms of molecular structures are ruled out. If applied to the universe or world as a whole it means that in the far future all life and all activity will cease to exist.

Since the 1870s the question of the cosmic heat death had been discussed endlessly by scientists, philosophers, theologians, and social critics. Far from being merely a scientific question related to thermodynamics and cosmology, it was a central issue in the cultural, ideological, and political struggle which in a German context is known as the *Kulturkampf*. The heat death was not only controversial because it predicted an end of the living universe but also because it suggested a cosmic beginning, a kind of creation [Kragh 2008]. After all, if the entropy increases continually, it seems to imply that it cannot have increased during an infinite span of time, for then we would live in a high-entropic world without any order at all. A beginning of the world does not necessarily require a supernatural or divine creation, but to people of a Christian orientation this was a tempting apologetic inference and one actually suggested in the period [Kragh 2008, pp. 60-93].

Weinstein examined the heat death and its consequences in detail in another book on philosophy of science published a few years later. From the point of view of what he called "spiritual monism" he reviewed the possibilities that after all the ultimate heat death might not occur. For example, he speculated that perhaps the end of the material world would result in a complete transformation of matter into a uniform sea of ether which might subsequently, by means of an as yet unknown mechanism, re-condense into particles of matter. Speculations of this kind were not quite new, but as an expert in thermodynamics Weinstein [1911] realized that they had no foundation in physics and hence that the entropic end of the world was probably a reality of the future. "All of life is a fight against the entropic death," he declared, and it was a fight that life, in so far it is a manifestation of matter, was doomed to lose.

And yet, as seen from Weinstein's spiritual and subjectivist point of view, the end of organic life might not be the end of the living universe. He suggested that "psychic energy" was even more primary than heat and ether, and that a maximum-entropy world would ultimately consist of this form of energy unknown to conventional physics. "That would correspond to the Indians' thoughts of Brahma and Buddha's nirvana," he commented, adding the rhetorical question, "Would we call such an end in an absolutely spiritual being death?" According to Weinstein [1911, p. 272], this kind of cosmic end was "certainly not death, but probably a dreamless sleep from which there is no awakening." Elsewhere in his book (p. 248) he reasoned that if the world comes to an end, it must have a beginning:

> The world cannot emerge by itself from the entropic death. If the world, understood as matter, substance or energy, is finite, … the entropic death must occur within a finite time. But then the world and its processes in particular, must also have begun a finite time ago. This cannot have happened by itself, … [and] a supernatural cause must consequently have been active. If one is forced to admit such a cause in the beginning, one can also let it govern the end, so that a beginning follows the end, and so on in all eternity.

In other words, if a divine being is called into action to create the world, why not let it perform the cosmic miracle more than once? Why not an infinite number of times? In yet another book Weinstein [1913a, pp. 102-118] returned to the problem of the heat death and its corollary in the form of entropic creation. He critically reviewed contemporary attempts to avoid the heat death by means of hypotheses of anti-entropic cosmic mechanisms, but found none of these attempts to be satisfactory.

Notice that Weinstein's argument rested on the assumption that the universe is spatially finite. In the discussion of the heat death and the origin of the universe during the late nineteenth century it was often taken for granted that the entropy law only applies to a finite universe. Weinstein apparently agreed, stating that only in such a universe would a cosmic beginning be inevitable. But is the universe actually finite in space and matter? Although Weinstein readily admitted that there was no scientifically valid answer to the question, he argued that an infinite universe was less probable than a finite one. Consequently, a cosmic beginning was more probable than an eternal past.

## 6. Pioneer of pandeism

Weinstein did not accept the theistic version of the entropic creation argument, but he agreed that the beginning of the universe was a problem inaccessible to conventional science just as the end of the universe was. "As far as I can see, only Spinozist pantheism, among all philosophies, can lead to a satisfactory solution," he wrote. And yet Weinstein did not fully subscribe to either Spinozism or pantheism.

"If only differing by a letter (d instead of t), we distinguish fundamentally between *pandeism* and pantheism." Thus wrote Weinstein [1910] in a book in which he discussed from a historical perspective the idea of pandeism, the doctrine that although the world is divinely created there is no God beyond or except the world itself. In a sense, by his act of creation God transformed into the world leaving nothing behind. As regards the term "pandeism" it became better known with Weinstein's *Welt- und Lebensanschauungen*, but he did not coin the word, which probably dates from 1859. The book attracted some attention in theological and philosophical circles, with a critical theological review appearing in [Kirn 1910].

According to Weinstein [1910, pp. 227-228], while pantheism was a metaphysical view, pandeism shared with deism that it belonged to the realm of religion. He explained: "The entire world is filled with mental deities [*Seelengottheiten*] and these deities have coalesced into a single … cosmic mental deity which permeates everything and has no meaning outside the world." There are certain similarities between the view of pandeism and the one of so-called panentheism as it appears in the process theology inspired by the metaphysical theories of A. N. Whitehead and Charles Hartshorne. But given that panentheism operates with an interacting God-world dualism, while at the same time it retains the distinct identities of both God and the world, there are also important differences. There is no such dualism in pandeism, which on the other hand shares with pantheism the monist view that the deity and the world are one and the same.

The idea of filling the universe with a kind of divine psychic or mental energy was not in itself very original as it was known from earlier idealistic philosophers such as J. G. Fichte and G. W. F. Hegel. In 1910 Felix Auerbach, a physics professor at the University of Jena, argued that there exists a vital "ectropy" in the universe able to counter the growth of entropy and thus to secure mind and life in an indeterminate future [Auerbach 1910]. In a vaguely similar way the British engineer and physicist Georg W. de Tunzelmann [1910] suggested the same year the existence of an all-pervading mind, which he likened to an entity even more refined and even more fundamental than the ether. Weinstein was aware of Auerbach's work but not, apparently, of Tunzelmann's.

In his *Welt- und Lebensanschauungen* Weinstein discussed extensively earlier philosophers and theologians whose views included elements of pandeism, trying to trace the concept as far back in time as possible. He paid particular attention to Giordano Bruno as an early pandeist. It is perhaps for his comprehensive investigation of the historical roots of pandeism, and of religious thoughts generally, that Weinstein is best known today. He has a place in the history of ideas more so than in the history of physics.

**Acknowledgments**. I thank Evgeni Starikov for information about Weinstein, and Christian Joas for help with translating passages in Weinstein's papers.